\documentclass[prd,tightenlines,nofootinbib,showpacs,preprintnumbers,superscriptaddress, twocolumn]{revtex4-1}
\usepackage{amsfonts,amsmath,amssymb,amsthm,bbm,hyperref}
\usepackage{graphicx}
\usepackage{color}

\newcommand{\be}{\begin{equation}}
\newcommand{\ee}{\end{equation}}
\newcommand{\beq}{\begin{eqnarray}}
\newcommand{\eeq}{\end{eqnarray}}
\newcommand{\nn}{\nonumber}

\DeclareMathOperator\cotanh{cotanh}

\begin{document}

\title{Invariant vacuum}
\author{Salvador Robles-P\'{e}rez}
\affiliation{Estaci\'{o}n Ecol\'{o}gica de Biocosmolog\'{\i}a, Pedro de Alvarado 14, 06411 Medell\'{\i}n, Spain.}
\affiliation{Instituto de  F\'{\i}sica Fundamental, CSIC, Serrano 121, 28006 Madrid, Spain}

\date{\today}

\begin{abstract}
We apply the Lewis--Riesenfeld invariant method for the harmonic oscillator with time dependent mass and frequency to the modes of a charged scalar field that propagates in a curved, homogeneous and isotropic spacetime. We recover the Bunch-Davies vacuum in the case of a flat DeSitter spacetime, the equivalent one in the case of a closed DeSitter spacetime and the invariant vacuum in a curved spacetime that evolves adiabatically. In the three cases, it is computed the thermodynamical magnitudes of entanglement between the modes of the particles and antiparticles of the invariant vacuum, and the modification of the Friedmann equation caused by the existence of the energy density of entanglement. The amplitude of the vacuum fluctuations are also computed.
\end{abstract}

\maketitle


\section{Introduction}

All the machinery of a quantum field theory is ultimately rooted on the definition of the vacuum state. Once this is defined a Fock space can be generated from the number eigenstates of the corresponding representation and the general quantum state of the field can be written as a vector of such  space. The field can then be interpreted as composed of many particles propagating along the spacetime. 

However, the definition of the vacuum state and the associated definition of particle cannot be always unambiguously stated in a curved spacetime. The most appropriate definition of the vacuum state in a local region of the spacetime may not correspond to the vacuum state in another local region, and that may lead to the creation of particles \cite{Parker1968, Parker1969, Parker2017, Mostepanenko1974, Mamaev1976, Fulling1979}. The question is then which vacuum state has to be selected from the set of possible vacuum states, with a twofold consideration: which quantum representation can determine the appropriate boundary condition for the field; and, which one can represent the observable particles.

A customary approach \cite{Birrell1982, Mukhanov2007} is to define the vacuum state in an "IN" and "OUT" regions that asymptotically behave like Minkowski spacetime, where the vacuum state is therefore well defined. The corresponding "IN" vacuum is assumed to supply the initial boundary condition for the field and the "OUT" vacuum is expected to define the kind of measurable particles. Generally, the result is that the initial vacuum state turns out to be full of particles of the "OUT" representation. A problem with this approach is  that it is not always possible to find in a curved spacetime two asymptotically flat regions where to define these vacuum states. That might wrongly induce us to think that a well defined vacuum state cannot be then given.

In this paper we shall adopt a different point of view. On the one hand, one would expect that the appropriate boundary condition for a cosmological field should be global, i.e. not tied to a local initial state, and such that the field should remain in the same state along the entire evolution of the field if no external force is present. In that case the state of the field should be invariant under time evolution. Furthermore, in cosmology there is no external element to the universe\footnote{We are not considering a multiverse scenario here. If that would be the case the same would apply to the multiverse as a whole instead of a single universe.} so in particular, one would expect the field to stay in the ground state or the state of minimal excitation of some invariant representation.

In most cases of interest the wave equation of the field modes in a curved spacetime turns out to be the wave equation of a harmonic oscillator with time dependent mass and frequency. Then, we can apply the method of the invariants of the harmonic oscillator, developed by Lewis--Riesenfield  \cite{Lewis1968, Lewis1969} and others \cite{Leach1983, Pedrosa1987, Dantas1992, Kanasugui1995,  Song2000, Vergel2009}, to find the invariant representation of the field modes. The important property of the invariant representation is that the associated number operator turns out to be a constant of motion. It means that once the field is in a given quantum superposition of the number eigenstates of the invariant representation it remains in the same state along the entire evolution of the field. In particular, if the field is in the vacuum state of the invariant representation at a given moment of time it will remain in the same vacuum state along the entire evolution of the field.

Then, we shall assume that the field is in the vacuum state of the invariant representation. Furthermore, instead of imposing an \emph{initial} condition on the state of the field at some given time $t_0$, we shall impose the \emph{boundary} condition that the largest modes of the field must be the positive frequency modes of a field that propagates in a Minkowski spacetime. This is a boundary condition that is ultimately rooted in the equivalence principle of the theory of relativity. For a sufficiently closed neighborhood, the spacetime looks always like a flat spacetime and, therefore, the largest modes of the field must not feel the curvature of the spacetime. This boundary condition will fix the invariant representation to be used and, thus, it will fix the invariant vacuum state.

In terms of the invariant representation the invariant vacuum state will then represent the ground state along the entire evolution of the field. However, in terms of the number states of any other representation the vacuum state of the invariant representation may contain particles. Let us notice that the concept of particle is a local concept that is based on the definition of the particle detector and, thus, the number of detected particles is an observer-dependent quantity. In particular, for an observer that is making measurements in a local region of the spacetime, the most appropriate representation of the vacuum seems to be the vacuum of instantaneous Hamiltonian diagonalization \cite{Mukhanov2007}, which represents the state of minimal excitation at a given moment of time. More concretely, an actual detector will only detect  particles with wavelength smaller than the characteristic length of the detector. We shall then show that such a detector will in practice detect no particles in a small local region of the spacetime because, as a consequence of the boundary condition, the field modes remain there in the vacuum state along the entire evolution of the field. However, on cosmological grounds, the invariant vacuum turns out to be full of particle-antiparticle pairs of the diagonal representation, which are created in entangled states. We can then analyze the quantum state of each component of the entangled pair and their evolution separately.

The paper is outlined as follows. In Sect. II we briefly review the customary procedure of canonical quantization of a charged scalar field. In Sect. III we obtain the invariant representation of the associated Hamiltonian and define the invariant vacuum state. In Sect. IV we apply the results to the case of a DeSitter spacetime and in Sect. V the same is done for a homogeneous and isotropic spacetime that evolves adiabatically. Finally, we summarize and draw some conclusions in Sect. VI.

\section{Field quantization}

Let us briefly summarize the standard procedure of canonical quantization for a charged scalar field, $\phi(x) = \phi(\textbf{x}, t)$, by starting from the action integral
\be\label{ACT01}
S = \int d t  d^3\textbf{x} \  \mathcal{L} = \int d t \ L ,
\ee
with the Lagrangian density $\mathcal{L}$ given by \cite{Bronnikov1968, Chernikov1968, Mamaev1976, Birrell1982}
\be\label{LAG01}
\mathcal{L}(x) =  \sqrt{- g} \left( g^{\mu \nu} \partial_\mu\phi \partial_\nu\phi^* - \left( m^2 + \xi R(x) \right) \phi(x) \phi^*(x)  \right) ,
\ee
where $m$ is the mass of the  field and $g_{\mu\nu}$ is the metric tensor, with $g \equiv \text{det}(g_{\mu\nu})$. The coupling between the scalar field and the gravitational field is represented by the term $\xi R \phi^2$, where $R(x)$ is the Ricci scalar. The value $\xi = 0$ corresponds to the so-called minimal coupling and the value $\xi = \frac{1}{6}$ corresponds to the conformal coupling. Unless otherwise indicated, we shall assume minimal coupling ($\xi = 0$)  but a similar procedure can be followed with any other value of $\xi$. The variational principle of the action (\ref{ACT01}) yields the  field equation
\be\label{FEq01}
\left( \Box_x + m^2 + \xi R(x) \right) \phi(x) = 0 ,
\ee
where the d'Alembertian operator $\Box_x$ is given by \cite{Birrell1982}
\be
\Box_x \phi = g^{\mu\nu} \nabla_\mu \nabla_\nu\phi = \frac{1}{\sqrt{-g}} \partial_\mu \left( \sqrt{-g} g^{\mu\nu} \partial_\nu \phi\right) .
\ee
In particular, let us consider a homogeneous and isotropic spacetime with metric element given by 
\be\label{ME01}
ds^2 =  d t^2 - a^2 \, dl^2 , 
\ee
where, $a=a(t)$ is the scale factor and $dl^2 = h_{i j } dx^i dx^j$, is the metric element of the three dimensional space with the constant curvature $\kappa = 0, \pm 1$. It is customary to work in conformal time $\eta$, and to scale the scalar field according to, $\phi = a^{-1} \chi$. In that case, the modes of the field $\chi$ satisfy  the wave equation of a harmonic oscillator with constant mass and time dependent frequency. However, we shall work in cosmic time $t$ and retain the charged scalar field $\phi(\textbf{x}, t)$ for at least for three reasons: i) the scaling is unnecessary for obtaining the invariant representation of the scalar field $\phi(x)$; ii) unlike in the wave equation of $\chi$, the frequency of the wave equation of $\phi$ is always real, so we shall avoid imaginary values of the frequency of the modes; and, iii) the invariant representation of any two field variables is the same provided that they are related by a canonical transformation, i.e. the invariant representation of the field $\chi(x)$ is also the invariant representation of the field $\phi(x)$, so the vacuum state of the invariant representation is the same for both fields. 

The isotropy of the spacetime described by the metric (\ref{ME01}) allows us to expand the field in Fourier modes
\be\label{FE01}
\phi(\textbf{x}, t) = \int d\mu(k) \psi_\textbf{k}(\textbf{x}) \phi_\textbf{k}(t)  ,
\ee
where $\psi_\textbf{k}$ are the eigenfunctions of the three-dimensional Laplacian, 
\be\label{SH01}
\Delta^{(3)}\psi_\textbf{k}(\textbf{x}) = - (k^2 - \kappa) \psi_\textbf{k}(\textbf{x}) , 
\ee
and, $k = |\textbf{k}|$ with $\textbf{k} = \{ k_x,k_y,k_z \}$ with $-\infty < k_i < \infty$ in the flat case,  or just $k$ in $\textbf{k} = \{ k, l, m \}$ with $0 < k < \infty$, $l = 0,1,2,\ldots$ in the open case, $k = 1,2,\ldots$ and $l = 0, 1,\ldots, k-1$ in the closed case, with $- l \leq m \leq l$ in both cases, and $d\mu(k)$ is the measure of the Fourier space (see Refs. \cite{Mostepanenko1974, Mamaev1976, Birrell1982} for the details). With (\ref{FE01}) and (\ref{SH01}), integrating by parts and using the orthogonality properties of the functions $\psi_\textbf{k}(\textbf{x})$ \cite{Birrell1982}, the Lagrangian in (\ref{ACT01}) turns out to be
\be\label{LAG03}
L = \int d\mu(k) M(t) \left\{ \dot{\phi}_\textbf{k}  \dot{\phi^*}_\textbf{k} - \omega_k^2(t) \, \phi_\textbf{k} \phi^*_\textbf{k} \right\} , 
\ee
where, $M(t) = a^3(t)$, 
\be\label{OM01}
\omega^2_k(t) = \frac{k^2- \kappa}{a^2} + m^2  + \xi R    .
\ee
The Lagrangian (\ref{LAG03}) is the Lagrangian of a set of harmonic oscillators with time dependent mass and frequency. Let us now proceed to quantize the field modes by writing \cite{Mamaev1976, Birrell1982, Mukhanov2007}
\be\label{FE02}
\phi_\textbf{k}(t) = \frac{1}{\sqrt{2}} \left( v_k(t) a_\textbf{k} + (- 1)^{\kappa m} v^*_k(t) b_{-\textbf{k}}^\dag \right) ,
\ee
where, $-\textbf{k} = \{-k_x, -k_y, -k_z\}$, in the flat case and, $-\textbf{k} = \{k, l, -m\}$ in the open and closed cases and, $\psi_\textbf{k}^* = (-1)^{\kappa m} \psi_{-\textbf{k}}$, for $\kappa = 0, \pm1$. In (\ref{FE02}), $a_\textbf{k}^\dag$ and $a_\textbf{k}$ are constant operators that describe the creation and annihilation operators of particles and $b_\textbf{k}^\dag$ and $b_\textbf{k}$ are those for antiparticles. They obey the standard commutation relations
\beq
[a_\textbf{k}, a_{\textbf{k}'}^\dag] &=& \delta(\textbf{k}-\textbf{k}')   ,  [a_\textbf{k}, a_{\textbf{k}'}] = [a_\textbf{k}^\dag, a_{\textbf{k}'}^\dag] = 0 , \\
{[b_{\textbf{k}}, b_{\textbf{k}'}^\dag]} &=& {\delta(\textbf{k}-\textbf{k}')  ,   [b_\textbf{k}, b_{\textbf{k}'}] = [b_\textbf{k}^\dag, b_{\textbf{k}'}^\dag] = 0} ,
\eeq
and define the vacuum state, $|0_a 0_b\rangle = |0\rangle_a |0\rangle_b$, as usual by the relation
\be
a_\textbf{k} |0 \rangle_a = 0 \ , \ b_\textbf{k} |0\rangle_b = 0 ,
\ee
for all $\textbf{k}$. The field amplitudes,  $v_k(t)$ in (\ref{FE02}), satisfy then
\be\label{WE01}
\ddot{v}_k+ \frac{\dot{M}}{M} \dot{v}_k + \omega_k^2(\eta) v_k = 0 .
\ee
Because the time dependence of the mass and frequency of the harmonic oscillator (\ref{WE01}) the vacuum state defined at $t_0$ contains particles and antiparticles at any other moment of time $t_1$. Therefore, it does not represent the no particle state along the evolution of the scalar field.

\section{Invariant vacuum state}

\subsection{Classical description}

There is a quantum representation that can describe a non-particle state along the entire evolution of the scalar field. It is given by the invariant representation. We shall briefly sketch the general procedure developed in Refs. \cite{Lewis1968, Lewis1969, Leach1983, Pedrosa1987, Dantas1992, Kanasugui1995,  Song2000, Vergel2009}. Particularly, we shall closely follow the formulation given in Refs. \cite{Leach1983, Kanasugui1995}. Let us therefore consider the following point transformation
\be
\zeta_\textbf{k} = \frac{1}{\sigma} \phi_\textbf{k} ,
\ee
where $\sigma \equiv \sigma_k(t)$ is an auxiliary real function that satisfies the non linear equation
\be\label{SIGm01}
\ddot{\sigma} + \frac{\dot{M}}{M} \dot{\sigma} + \omega_k^2 \sigma = \frac{k^2}{M^2 \sigma^3} ,
\ee
with the frequency $\omega_k$ being given by (\ref{OM01}). Let us here notice that a solution of (\ref{SIGm01}) can be generally  given by\footnote{For more general solutions of (\ref{SIGm01}) see Ref. \cite{Leach1983}.}
\be\label{SIGm02}
\sigma = \sqrt{\sigma_1^2 + \sigma_2^2} ,
\ee
where $\sigma_1$ and $\sigma_2$ are two real independent solutions of
\be\label{SIGm03}
\ddot{\sigma}_{1,2} + \frac{\dot{M}}{M} \dot{\sigma}_{1,2} + \omega_k^2 \sigma_{1,2} = 0 ,
\ee
with the normalization condition, $\sigma_1 \dot{\sigma}_2 - \sigma_2 \dot{\sigma}_1 = \frac{k}{M}$. Let us also perform the following change of time variable, $t \rightarrow \tau_k$, given by
\be\label{TAU01}
d\tau_k = \frac{1}{M \sigma_k^2} d t .
\ee
Then, the action (\ref{ACT01}) with the Lagrangian (\ref{LAG03}) transforms into
\be\label{ACT03}
S = \int d\mu(k) S_\textbf{k} ,
\ee
where
\be
S_\textbf{k} =  \int d\tau_k \left\{ \frac{d\zeta_\textbf{k}}{d \tau_k} \frac{d\zeta^*_\textbf{k}}{d \tau_k} - k^2 \zeta_\textbf{k}^* \zeta_\textbf{k} \right\} ,
\ee
is the action of a harmonic oscillator with constant frequency $k$. The action (\ref{ACT03}) is the sum of the actions of a set of uncoupled harmonic oscillators, each one evolving however with a different time variable, $\tau_k$. The  momenta conjugated to $\zeta_\textbf{k}$ and $\zeta^*_\textbf{k}$ are
\be
\tilde{\pi}_\textbf{k} = \frac{d\zeta^*_\textbf{k}}{d \tau_k}  \ , \ \tilde{\pi}^*_\textbf{k} = \frac{d\zeta_\textbf{k}}{d \tau_k} ,
\ee
and the corresponding Hamiltonian reads
\be\label{HAM04}
\tilde{H}_\textbf{k} = \tilde{\pi}_\textbf{k} \tilde{\pi}_\textbf{k}^* + k^2 \zeta_\textbf{k} \zeta_\textbf{k}^* .
\ee
The wave equation for the field $\zeta_\textbf{k}(\tau_k)$ is
\be
\frac{d^2 \zeta_\textbf{k}}{d\tau_k^2} + k^2 \zeta_\textbf{k}^2 = 0 ,
\ee
with normalized solutions given by
\be\label{IM01}
\zeta_\textbf{k} = \frac{1}{\sqrt{k }} e^{- i k \tau_k} ,
\ee
which is positive frequency with respect to $\tau_k$. Then, the corresponding solutions of the original field modes are 
\be\label{WE05}
\phi_\textbf{k} = \frac{\sigma}{\sqrt{k}} e^{- i k \tau_k}  =  \frac{\sigma}{\sqrt{k }} e^{- i k \int \frac{1}{M \sigma^2} dt} .
\ee
The invariant value of the field $\phi$ relies then in the computation of the auxiliary function $\sigma$. In order to fix the value of $\sigma$ we must impose a boundary condition. For this, one has to realize that in terms of the rescaled field, $\chi = a \phi$, and in conformal time, $\eta = \int\frac{dt}{a}$, the wave equation (\ref{WE01}) becomes, in the limit of large modes of the field, the customary equation of a harmonic oscillator with unit mass and constant frequency $k$, i.e.
\be\label{WEAS01}
\chi_k'' + k^2 \chi_k = 0 ,
\ee
where the prime denotes derivative with respect to the conformal time. The positive frequency solutions of (\ref{WEAS01}) are
\be\label{PFS01}
\chi_k(\eta) = \frac{1}{\sqrt{k}} e^{-i k |\eta|} . 
\ee
Then, in order for the field modes (\ref{WE05}) to be the modes associated to the positive frequency solutions (\ref{PFS01}) we have to impose the boundary condition 
\be\label{BC01}
\sigma = a^{-1} \ , \ (\Rightarrow \tau_k = \eta) ,
\ee 
in the limit of large modes, $k \gg 1$. The normalization condition given after (\ref{SIGm03}) and the boundary condition (\ref{BC01}) fix the invariant representation to be used and, thus, they fix the invariant vacuum state.

Let us finally point out that the transformation
\be
(\phi_\textbf{k}, \phi^*_\textbf{k}, p_{\phi_\textbf{k}}, p_{\phi_\textbf{k}}^*; t) \rightarrow (\zeta_\textbf{k}, \zeta^*_\textbf{k}, {\pi}_\textbf{k}, {\pi}_\textbf{k}^*; \tau_k) ,
\ee
is a canonical transformation given by
\beq\label{CT01a}
\zeta_\textbf{k} = \frac{1}{\sigma} \phi_\textbf{k} \ &,& \ {\pi}_\textbf{k} = \sigma p_{\phi_\textbf{k}} - M \dot{\sigma} \phi_\textbf{k}^* , \\ \label{CT01b}
\zeta_\textbf{k}^* = \frac{1}{\sigma} \phi_\textbf{k}^* \ &,& \ {\pi}_\textbf{k}^* = \sigma p_{\phi_\textbf{k}}^* - M \dot{\sigma} \phi_\textbf{k} ,
\eeq
which is generated by the following generating function \cite{Kanasugui1995}
\be
F_2(\phi_\textbf{k}, \phi_\textbf{k}^*, {\pi}_\textbf{k}, {\pi}^*_\textbf{k}) = \frac{1}{\sigma} \left( \phi_\textbf{k} {\pi}_\textbf{k} + \phi_\textbf{k}^* {\pi}_\textbf{k}^* \right)  + \frac{M \dot{\sigma}}{\sigma} \phi_\textbf{k} \phi_\textbf{k}^* ,
\ee
through the relations \cite{Kanasugui1995}
\beq
& & p_{\phi_\textbf{k}} = \frac{\partial}{\partial \phi_\textbf{k}}  F_2 \ , \ p_{\phi_\textbf{k}}^* = \frac{\partial}{\partial \phi^*_\textbf{k}}  F_2  \\
& & \zeta_\textbf{k} = \frac{\partial }{\partial {\pi}_\textbf{k}}  F_2 \ , \ \zeta^*_\textbf{k} = \frac{\partial }{\partial {\pi}^*_\textbf{k}}  F_2 \\
& &H(\zeta_\textbf{k}, {\pi}_\textbf{k}, \zeta^*_\textbf{k}, {\pi}^*_\textbf{k}) \dot{\tau}_k = H(\phi_\textbf{k}, p_{\phi_\textbf{k}} , \phi^*_\textbf{k}, p_{\phi_\textbf{k}} ^*; t) + \partial_t F_2
\eeq

\subsection{Invariant creation and annihilation operators}

The Hamiltonian (\ref{HAM04}) is the Hamiltonian of a harmonic oscillator with unit mass and constant frequency given by $k$. Thus, the creation and annihilation operators defined in terms of the field $\zeta_\textbf{k}$ are invariant under the time evolution. Therefore, the annihilation operator of the invariant representation defines a vacuum state that is stable along the entire evolution of the scalar field. The invariant representation of particles and antiparticles, $\tilde{a}_\textbf{k}$, $\tilde{a}_\textbf{k}^\dag$ and $\tilde{b}_{-\textbf{k}}$, $\tilde{b}_{-\textbf{k}}^\dag$, respectively, is defined in terms of the invariant field and its conjugated momenta as usual, by
\beq\label{IRF01a}
\zeta_\textbf{k} = \frac{1}{\sqrt{2 k}} (\tilde{a}_\textbf{k} + \tilde{b}^\dag_{-\textbf{k}})  \ &,& \ {\pi}^*_\textbf{k} = - i \sqrt{\frac{k}{2}} (\tilde{a}_\textbf{k} - \tilde{b}^\dag_{-\textbf{k}})  , \\ \label{IRF01b}
\zeta_\textbf{k}^* = \frac{1}{\sqrt{2k}} (\tilde{b}_{-\textbf{k}} + \tilde{a}^\dag_\textbf{k} ) \ &,& \ {\pi}_\textbf{k} = -i \sqrt{\frac{k}{2}} (\tilde{b}_{-\textbf{k}} - \tilde{a}^\dag_\textbf{k} ) ,
\eeq
in terms of which the Hamiltonian (\ref{HAM04}) reads
\be
H_\textbf{k} = k \left( \tilde{a}_\textbf{k} ^\dag \tilde{a}_\textbf{k} + \tilde{b}^\dag_{-\textbf{k}} \tilde{b}_{-\textbf{k}} +  1 \right) .
\ee
Using (\ref{CT01a}-\ref{CT01b}) and the inverse relation of (\ref{IRF01a}-\ref{IRF01b}) we can express the invariant representation in terms of the original field modes $\phi_\textbf{k}$ and the conjugated momentum $p_\textbf{k}$. It yields (see the analogy with the invariant representation given in Refs.  \cite{Lewis1969, RP2010})
\beq\label{IR02a}
\tilde{a}_\textbf{k} &=& \sqrt{\frac{k}{2}} \left(  \frac{1}{\sigma} \phi_\textbf{k} + \frac{i}{k} ( \sigma p_{\phi_\textbf{k}}^*  - M \dot{\sigma} \phi_\textbf{k}  )  \right) , \\ \label{IR02b}
\tilde{b}_{-\textbf{k}} &=& \sqrt{\frac{k}{2}} \left(  \frac{1}{\sigma} \phi_\textbf{k}^* + \frac{i}{k} ( \sigma p_{\phi_\textbf{k}}  - M \dot{\sigma} \phi^*_\textbf{k} )  \right) .
\eeq
The important property of the invariant representation is that the eigenstates of the number operators of particles and antiparticles, $\tilde{N}_\textbf{k} ^a \equiv \tilde{a}_\textbf{k} ^\dag \tilde{a}_\textbf{k}$ and $\tilde{N}_\textbf{k} ^b \equiv \tilde{b}_\textbf{k} ^\dag \tilde{b}_\textbf{k}$, respectively, are stable along the entire evolution of the scalar field, because
\be\label{Nt01}
\frac{d \tilde{N}_\textbf{k}}{dt} = \dot{\tau}  \frac{d \tilde{N}_\textbf{k}}{d\tau} =  - i \dot{\tau}  [\tilde{N}_\textbf{k}, H_\textbf{k}]  = 0 .
\ee
It means that once the field is in a given eigenstate of the invariant number operator, or more generally in a quantum superposition of number eigenstates, it remains in the same state along the entire evolution of the spacetime. In particular, the vacuum state of the invariant representation, defined as $|\tilde{0}_a \tilde{0}_b\rangle = |\tilde{0}\rangle_a |\tilde{0}\rangle_b$, with
\be
\tilde{a}_\textbf{k} |\tilde{0} \rangle_a = 0 \ , \ \tilde{b}_\textbf{k} |\tilde{0}\rangle_b = 0 , \ \forall \ \textbf{k} ,
\ee
describes the no particle state along the entire evolution of the field irrespective of whether there is or not an asymptotically flat region of the spacetime. It is therefore a stable definition for the vacuum state and an appropriate representation to provide a global, observer-independent boundary condition for the state of the field.

\subsection{Relation with the diagonal representation}

At a given moment of time $t_0$, however, and for small changes around $t_0$ the representation that describes  instantaneously the ground state of the Hamiltonian is the diagonal representation, ${c}_\textbf{k}$, ${c}_\textbf{k}^\dag$ and ${d}_{-\textbf{k}}$, ${d}_{-\textbf{k}}^\dag$, defined as
\beq\label{DR01a}
\phi_\textbf{k} &=& \frac{1}{\sqrt{2 M \omega_k}} \left( {c}_\textbf{k} + {d}_{-\textbf{k}}^\dag \right) , \\ \label{DR01b}
p_{\phi_\textbf{k}}^* &=& - i \sqrt{\frac{M \omega_k}{2}} \left( {c}_\textbf{k} - {d}_{-\textbf{k}}^\dag \right) .
\eeq 
The instantaneous diagonal representation of the Hamiltonian at a given moment of time cannot define a stable vacuum state of the field because it entails the continuous generation of particles detected by a local particle detector at any other moment of time \cite{Fulling1979}. This can easily be seen by noting that, because the time dependence of $M$ and $\omega_k$ in (\ref{DR01a}-\ref{DR01b}), two different representations, $c_0 \equiv c_\textbf{k}(t_0)$ and $d_0 \equiv d_{-\textbf{k}}(t_0)$, and, $c_1 \equiv c_\textbf{k}(t_1)$ and $d_1 \equiv d_{-\textbf{k}}(t_1)$, at two given moments of time $t_0$ and $t_1$ are related by the Bogolyubov transformation
\beq\label{DR02a}
c_1 &=& \mu_0 \, {c}_0 - \nu^*_0 \, {d}_0^\dag  , \\ \label{DR02b}
d_1 &=& \mu_0 \, d_0 - \nu_0^* \, c_0^\dag ,
\eeq
where
\beq\label{MU001}
\mu_0 &=& \frac{1}{2} \left( \sqrt{\frac{M_1 \omega_1}{M_0 \omega_0}} +  \sqrt{\frac{M_0 \omega_0}{M _1\omega_1}}  \right) , \\ \label{NU001}
\nu_0= &=&  \frac{1}{2} \left( \sqrt{\frac{M_1 \omega_1}{M_0 \omega_0}} -  \sqrt{\frac{M_0 \omega_0}{M _1\omega_1}}  \right) ,
\eeq
with, $|\mu_0|^2 - |\nu_0|^2 = 1$, and, $M_{0,1} \equiv M(t_{0,1})$ and $\omega_{0,1} \equiv \omega_k(t_{0,1})$. In the limit of large modes the particles measured in a local region of the space at time $t_1$ would then be given, in an expanding universe, by
\be
N(t_1) = |\nu_0|^2 \approx \frac{M_1 \omega_1}{4 M_0 \omega_0} \sim \left( \frac{a(t_1)}{a(t_0)} \right)^2 \, , \forall {k} \gg 1 .
\ee
It means that a local particle detector would detect a large amount of particles in a large expanding universe like ours. It does not seem to be therefore a consistent boundary condition to impose that the field has to be in the vacuum state of the diagonal representation at a given initial time $t_0$.

A more appropriate boundary condition seems to be imposing that the field is in the vacuum state\footnote{Or generally speaking, in a linear combination of number states.} of the invariant representation. First, because of (\ref{Nt01}), the invariant vacuum state represents the no-particle state along the entire evolution of the field. Then, in terms of the invariant representation there is no particle production at all for all time \cite{Birrell1982}. However, we shall assume that the measurable particles are given, in a local region, by the number states of the instantaneous diagonal representation. Even though, we shall now show that a local detector will in practice detect no particles within a small region of the spacetime. Let us first notice that the invariant representation (\ref{IR02a}-\ref{IR02b}) can be related to the diagonal representation (\ref{DR01a}-\ref{DR01b}) through the Bogolyubov transformation
\beq\label{BT01a}
\tilde{a}_\textbf{k} &=& \mu(t) \, {c}_{\textbf{k}} - \nu^*(t) \, {d}_{-\textbf{k}}^\dag  , \\ \label{BT01b}
\tilde{b}_{-\textbf{k}} &=&   \mu(t) \, {d}_{-\textbf{k}} - \nu^*(t) \, {c}_{\textbf{k}}^\dag ,
\eeq
where
\beq\label{MU02}
\mu(t) &=& \frac{1}{2} \left( \sigma \sqrt{\frac{M \omega_k}{k}} + \frac{1}{\sigma} \sqrt{\frac{k}{M \omega_k}}  - i\dot{\sigma} \sqrt{\frac{M}{\omega_k k}}  \right) , \\ \label{NU02}
\nu(t) &=&\frac{1}{2} \left( \sigma \sqrt{\frac{M \omega_k}{k}} - \frac{1}{\sigma} \sqrt{\frac{k}{M \omega_k}}  - i\dot{\sigma} \sqrt{\frac{M}{\omega_k k}} \right) ,
\eeq
with, $|\mu|^2 - |\nu|^2 = 1$ for all time. In the limit of a large value of the mode $k$, i.e. within a small volume of the space, $\omega_k \sim \frac{k}{a}$ and $\sigma \sim a^{-1}$ (see (\ref{BC01})), and thus
\be\label{Nk01}
N_k = |\nu|^2 \rightarrow \frac{\dot{a}^2}{4 k^2} \sim \left( \frac{\lambda_\text{ph}}{H^{-1}} \right)^2 ,
\ee
where, $\lambda_\text{ph}= \frac{2 \pi a}{k}$, is the physical wavelength of the mode and, $H^{-1} = \frac{a}{\dot{a}}$, is the curvature radius at a given time. It can be easily seen from (\ref{Nk01}) that for sub-horizon modes, $\lambda_\text{ph} \ll H^{-1}$, the field does not feel the curvature of the spacetime and these modes remain in the vacuum state. A local particle detector of a practical length scale would measure then no particle at all within a small region of the space, irrespective of the moment of time. On cosmological grounds, however, there is a significant production of modes\footnote{These modes would be like \emph{global} particles in the sense that their associated wavelength are of order of the curvature radius.} but this is not surprising in an expanding universe whose evolution is determined, according to the Friedmann equation, by the matter content of the universe. The energy of the spacetime is negative and it balances the energy of the matter fields so the total energy is zero (see, for instance, Ref \cite{RP2017b}). Therefore, in an expanding universe the energy of the field is not conserved and it grows as the universe expands.

\subsection{Thermodynamical magnitudes of entanglement}

Let us now assume that the field is in the vacuum state of the invariant representation. It seems to be an appropriate boundary condition because it means that the field will remain in the same vacuum state along the entire evolution of the field, with a quantum state described by the density matrix
\be\label{RHO01}
\rho = | \tilde{0}_a \tilde{0}_b \rangle \langle \tilde{0}_a \tilde{0}_b | .
\ee
Using the Bogolyubov transformation (\ref{BT01a}-\ref{BT01b}) the vacuum state of the invariant representation can be written as \cite{Mukhanov2007}
\be
| \tilde{0}_a \tilde{0}_b \rangle = \prod_\textbf{k} \frac{1}{|\mu|} \left( \sum_{n=0}^\infty \left( \frac{\nu}{\mu} \right)^n | n_{c,\textbf{k}} n_{d,-\textbf{k}}\rangle \right) ,
\ee
where, $\mu \equiv \mu_k$ and $\nu \equiv \nu_k$, and
\be
|n_{c, \textbf{k}}\rangle = \frac{(c_\textbf{k}^\dag)^n}{\sqrt{n!}} |0_{c,\textbf{k}}\rangle \ , \ |n_{d,- \textbf{k}}\rangle = \frac{(d_{-\textbf{k}}^\dag)^n}{\sqrt{n!}} |0_{d,-\textbf{k}}\rangle ,
\ee
are the number states of the diagonal representation (\ref{DR01a}-\ref{DR01b}). It means that the vacuum state of the invariant representation is full of particle-antiparticle pairs created with opposite momenta in entangled states. Let us consider just the quantum state of the particles. The reduced density matrix that represents the quantum state of the particles alone can be obtained by tracing out from the density matrix (\ref{RHO01}) the degrees of freedom of the antiparticles. It typically yields \cite{RP2010, RP2017a, RP2017b}
\be\label{RHO02}
\rho_c = \text{Tr}_d {\rho}= \prod_\textbf{k} \frac{1}{Z_k} \sum_n e^{-\frac{1}{T_k} (n+\frac{1}{2})} | n_{c,\textbf{k}} \rangle \langle n_{c,\textbf{k}}| ,
\ee
where, $Z_k^{-1} = 2 \sinh\frac{1}{2T_k}$, with a specific temperature of entanglement \cite{RP2017a} given by
\be
T_k \equiv T_k(t) = \frac{1}{ \ln \frac{|\mu(t)|^2}{|\nu(t)|^2}} = \frac{1}{\ln\left( 1 + |\nu(t)|^{-2}\right)} .
\ee
The temperature of entanglement is a measure of the entanglement between the particles and antiparticles of the charged scalar field. Therefore, it is also a measure of the effects of the curvature of the spacetime. For a large value of the mode $k$, $T_k \rightarrow 0$, and there is thus no entanglement, as it is expected because the largest modes (or at the shortest distances) do not feel the curvature of the spacetime.

One can even define the thermodynamical magnitudes of entanglement associated to the quasi thermal state (\ref{RHO02}). They are given, for each mode, by \cite{RP2012}
\beq\label{Eent}
E_k(t) &=& \frac{\omega_k}{2} \cotanh\frac{1}{2 T_k} = \omega_k \left( N_k +\frac{1}{2}\right) , \\
Q_k(t) &=& \frac{\omega_k}{2} \cotanh\frac{1}{2 T_k} - \omega_k T_k \ln\sinh\frac{1}{2 T_k}, \\
W_k(t) &=& \omega_k T_k \ln\sinh\frac{1}{2 T_k} ,
\eeq
where, $N_k \equiv |\nu|^2$. The first principle of thermodynamics, $E_k(t) = Q_k(t) + W_k(t)$, is satisfied for all modes $k$ individually, and the energy densities that correspond to $E_n$, $Q_n$, and $W_n$, are given by
\be\label{edensity}
\varepsilon_n = \frac{E_n}{V} \ , \ q_n = \frac{Q_n}{V} \ , \ w_n = \frac{W_n}{V} ,
\ee
with, $V = a^3(t)$. The entropy of entanglement \cite{Horodecki2009, RP2012} can also be easily obtained from the von Neumann formula
\be
S(\rho) = - \rm{Tr}\left( \rho \ln \rho \right) ,
\ee
with $\rho$ given by (\ref{RHO02}). It yields \cite{RP2012}
\begin{equation}
\label{eq68}
S_\text{ent}(a) = |\mu|^2 \, \ln |\mu|^2 - |\nu|^2 \, \ln |\nu|^2.
\end{equation}

\section{DeSitter spacetime}

\subsection{Flat DeSitter spacetime}

Let us now consider a flat DeSitter spacetime described by the metric element (\ref{ME01}) and a scale factor given by
\be\label{SF01fds}
a(t) =  \frac{1}{H} e^{H t} ,
\ee
where, $-\infty < t < \infty$, and $\Lambda \equiv H^2$ is the cosmological constant. The invariant representation is given by (\ref{IR02a}-\ref{IR02b}) with the function $\sigma$ being given by (\ref{SIGm02}) with $\sigma_1$ and $\sigma_2$ satisfying
\be
\ddot{\sigma}_{1,2} + 3 H \dot{\sigma}_{1,2} + \left( H^2 k^2 e^{-2H t} + m^2\right) \sigma_{1,2} = 0 .
\ee
The two solutions that make $\sigma$ in (\ref{SIGm02}) satisfying the boundary condition (\ref{BC01}) are,
\beq\label{SIG1fds}
\sigma_1(t) &=& H \sqrt{\frac{\pi k}{2}} e^{-\frac{3 H}{2} t} \mathcal{J}_\mu(k e^{-Ht}) , \\ \label{SIG2fds}
\sigma_2(t) &=& H \sqrt{\frac{\pi k}{2}} e^{-\frac{3 H}{2} t} \mathcal{Y}_\mu(k e^{-Ht}) ,
\eeq
where
\be\label{MU01}
\mu = \sqrt{\frac{9}{4} - \frac{m^2}{H^2}} .
\ee
Let us notice that with the value of $\sigma$ given by (\ref{SIGm02}) with $\sigma_1$ and $\sigma_2$ given by (\ref{SIG1fds}-\ref{SIG2fds}), the invariant field modes (\ref{WE05}) are nothing more than the modes associated to the Bunch-Davies vacuum, as it was expected. By using the properties of the Bessel functions \cite{Abramovitz1972}, one can easily check that 
\be
\sigma = \sqrt{\sigma_1^2 + \sigma_2^2} = \frac{1}{a} \sqrt{\frac{\pi k}{2 H a}} |\mathcal{H}_\mu^{(2)}(x)| ,
\ee
where $\mathcal{H}_\mu^{(2)}(x)$ is the Hankel function of second kind and order $\mu$, with 
\be
x = k e^{- H t} = \frac{k}{Ha} = k |\eta| ,
\ee
where $\eta$ is the conformal time and, $\theta \equiv - k \tau_k$, is the phase of the Hankel function satisfying (see (9.2.21) of Ref. \cite{Abramovitz1972})
\be
|\mathcal{H}_\mu^{(2)}(x)|^2 \frac{d\theta}{d x} = \frac{2}{\pi x} .
\ee
Therefore,
\be
\phi_\textbf{k} = \frac{\sigma}{\sqrt{k}} e^{- i k |\tau_k|} = \frac{1}{a} \sqrt{\frac{\pi |\eta|}{2}} \mathcal{H}_\mu^{(2)}(k|\eta|)  .
\ee
It means that the Bunch-Davies vacuum corresponds to the invariant vacuum in the sense of the Lewis-Riesenfeld formulation too, as it was expected \cite{Kim1999}.  But it also means that we can find the values of $\sigma$ and $\tau_k$ in (\ref{SIGm01}) and (\ref{TAU01}), respectively, by computing the modulus and phase of the invariant wave function of the field propagating in a more general, curved spacetime.

In terms of the diagonal representation, the number of particles of the field in the flat DeSitter spacetime,
\be\label{Nflat}
N_a = |\nu |^2 \approx \frac{9 H^2}{16 k_\text{ph}^2} \sim \left( \frac{L_\text{ph}}{H^{-1}} \right)^2  ,
\ee
on a given physical scale, $k_\text{ph} = k/a$, does not depend on time (in the limit $k\gg 1$), and it is negligible for a practical detector of human length scale. For large values of the scale factor the energy density of the particles is given by
\be\label{ED01fds}
\varepsilon \approx \int_0^{k_\text{m}} dk \ k^2 \varepsilon_k \propto \frac{9 H^2 k_\text{m}^2}{16 a^2} .
\ee
It is therefore a function that decreases exponentially in time for an evolved universe like ours.

Finally, the amplitude of fluctuations of the field can be easily obtained from
\be\label{AF01fds}
\delta\phi_\textbf{k} = \frac{k^\frac{3}{2}}{2 \pi} \Delta\phi_\textbf{k} ,
\ee
with
\be
(\Delta\phi_\textbf{k})^2 = \langle |\phi_\textbf{k}|^2 \rangle - |\langle \phi_\textbf{k} \rangle|^2  = \frac{\sigma^2}{2 k} ,
\ee
where the expected values are computed in the vacuum state of the invariant representation and $\phi_\textbf{k}$ can be obtained from the inverse relation of (\ref{IR02a}-\ref{IR02b}). It gives the standard expressions for the spectrum of fluctuations \cite{Mukhanov2007}
\be\label{SFfds}
\delta\phi(k_\text{ph}) = \frac{H}{4\sqrt{\pi}} \left( \frac{k_\text{ph}}{H} \right)^\frac{3}{2} \left( \mathcal{J}^2_\mu(\frac{k_\text{ph}}{H}) + \mathcal{Y}^2_\mu(\frac{k_\text{ph}}{H}) \right)^\frac{1}{2} .
\ee

\subsection{Closed DeSitter spacetime}

The relation between the Lewis--Riesenfeld formalism of the invariant modes and the customary formulation of the invariant wave function is not restricted to the flat DeSitter spacetime and it is indeed quite general. Let us notice that $\sigma$ and $\tau_k$ are nothing more the modulus and the phase of the wave function (\ref{WE05}), and that the equations (\ref{SIGm01}) and (\ref{TAU01}) are, respectively, the real and the complex parts of the wave equation (\ref{WE01}) for the modes (\ref{WE05}), i.e. inserting (\ref{WE05}) in (\ref{WE01}) one obtains (\ref{SIGm01}) and (\ref{TAU01}). Then, the customary solutions of the modes of a scalar field in a curved spacetime are recovered here by taking the appropriate solutions of $\sigma_1$ and $\sigma_2$. On the other hand, the values of $\sigma$ and $\tau_k$ can be obtained by computing the modulus and phase of the normalized solutions of the wave equation of the scalar field.

In the case of a closed DeSitter spacetime with metric element (\ref{ME01}) and a scale factor given by
\be\label{SF01cds}
a(t) = \frac{1}{H}  \cosh Ht ,
\ee
where, $-\infty < t < \infty $, the functions $\sigma_1(t)$ and $\sigma_2(t)$ satisfy (\ref{SIGm03}) with the frequency $\omega_k$ given by
\be\label{OM01cds}
\omega_k^2 = \frac{H^2 (k^2 - 1)}{\cosh^2Ht} + m^2 .
\ee
The solutions of (\ref{SIGm03}) with the frequency (\ref{OM01cds}) can be written in terms of the hypergeometric functions \cite{Gutzwiller1956, Chernikov1968, Mottola1985}, or equivalently in terms of Legendre functions \cite{Gutzwiller1956, Tagirov1973, Birrell1982}. As it is pointed out in Ref. \cite{Birrell1982}, if one follows the procedures used in Ref. \cite{Gutzwiller1956} (see also Refs. \cite{Mottola1985, Dowker1976}) and defines the IN and OUT vacuum states by taking the positive frequency solutions of (\ref{SIGm03}) in the asymptotic limits, $t \rightarrow \pm\infty$, one obtains an infinite particle production, irrespective of the value of $k$. It means that one should measure an infinite number of particles even in a small region of the space as the universe expands.

On the contrary, we are here imposing the boundary condition that the field is in the vacuum state of the invariant representation, in terms of which there is no particle production at all at any time because it represents the no particle state along the entire evolution of the field. Following Ref. \cite{Chernikov1968} (see also, Ref. \cite{Mottola1985}) we can express the solutions of (\ref{SIGm03}) in terms of the hypergeometric function as
\beq\nn
\chi_k = \frac{1}{k!} & & \sqrt{\Gamma(k+\frac{1}{2}-\mu) \Gamma(k+\frac{1}{2}+\mu)} e^{-i k \eta} \\  \label{CHI01}
& & \times F(\frac{1}{2}- \mu, \frac{1}{2}+\mu ; 1+k; \frac{1-i \tan\eta}{2}) ,
\eeq
where, $-\frac{\pi}{2} < \eta < \frac{\pi}{2}$, is the conformal time and, $F(a,b;c;z) = _2F_1(a,b;c;z)$, is the hypergeometric function. By taking into account the expansion of (\ref{CHI01}) in powers of $k$, given by \cite{Chernikov1968}
\be
\chi_k = \frac{1}{\sqrt{k}} e^{-i k \eta} \left( 1 + \mathcal{O}(k^{-1}) \right) ,
\ee
one can easily  check that the modes (\ref{CHI01}) reduce to  (\ref{PFS01}) in the limit of large modes ($k\rightarrow \infty$), so the modes (\ref{CHI01}) already satisfy our boundary condition. However, in order to give an explicit expression  of $\sigma$ it is more convenient to rewrite the modes (\ref{CHI01}) as \cite{Birrell1982, Tagirov1973} 
\be\label{BDM01}
\chi_k(\eta) = N_k \cos^\frac{1}{2} \eta \left( P_{k-\frac{1}{2}}^\mu(\sin\eta)  - \frac{2 i}{\pi }  Q_{k-\frac{1}{2}}^\mu(\sin\eta) \right) ,
\ee
where $P_\nu^\mu(x)$ and $Q_\nu^\mu(x)$ are the associated Legendre functions of first and second kind, respectively, of degree $\nu$ and order $\mu$, being $\mu$ given by (\ref{MU01}), and
\be
N_k = \left( \frac{\pi \, \Gamma(k + \frac{1}{2} - \mu)}{2 \, \Gamma(k+\frac{1}{2}+ \mu)} \right)^\frac{1}{2}  e^{i\frac{\mu \pi}{2}} .
\ee
Then, the appropriately normalized solutions of $\sigma_1$ and $\sigma_2$ in (\ref{SIGm03}) are given by
\beq\label{SIG1cds}
\sigma_1 &=&  \frac{ |N_k|  \sqrt{k}}{a} \sin^\frac{1}{2}\eta P_{k-\frac{1}{2}}^\mu(-\cos\eta) , \\ \label{SIG2cds}
\sigma_2 &=& -   \frac{2 |N_k| \sqrt{k}}{\pi \, a} \sin^\frac{1}{2}\eta Q_{k-\frac{1}{2}}^\mu(-\cos\eta) ,
\eeq
so that (see (\ref{WE05})), 
\be
\sigma = \sqrt{\sigma_1^2 + \sigma_2^2} = \frac{\sqrt{k}}{a} |\chi_\textbf{k}| .
\ee
It can also be checked that
\be
\tau_k = \frac{1}{k} \arctan\frac{2 \, Q_{k-\frac{1}{2}}^\mu(-\cos\eta) }{\pi \, P_{k-\frac{1}{2}}^\mu(-\cos\eta)} - \frac{ \mu \pi}{2} ,
\ee
satisfies (\ref{TAU01}).

The value of $\nu$ in (\ref{NU02}) turns out to be then
\be\label{NUkcds}
\nu(t) \approx - i (\mu - \frac{3}{2}) \frac{\dot{a}}{k} = - i (\mu - \frac{3}{2}) \frac{\sqrt{H^2 a^2 - 1}}{k} \rightarrow 0 ,
\ee
in the  limit, $k \rightarrow \infty$, for all time $t$,  so again the vacuum state of the invariant representation (\ref{IR02a}-\ref{IR02b}), with $\sigma$ given by (\ref{SIGm02}) with $\sigma_1$ and $\sigma_2$ given by (\ref{SIG1cds}-\ref{SIG2cds}), defines a stable adiabatic vacuum state. The energy density behaves similar to (\ref{ED01fds}) for large values of the scale factor, as it was expected.

The amplitude of fluctuations (\ref{AF01fds}) gives now,
\beq\nn
\delta\phi(k_\text{ph}) &\approx & \frac{H}{4\sqrt{\pi}} \left( \frac{k_\text{ph}}{H} \right)^\frac{3}{2} \sqrt{\frac{\Gamma(k+\frac{1}{2}-\mu)}{\Gamma(k+\frac{1}{2}+\mu)} } \\ & & \times \sqrt{  \left( P^\mu_{k-\frac{1}{2}}(z) \right)^2  + \frac{4}{\pi^2} \left( Q^\mu_{k-\frac{1}{2}}(z) \right)^2 } ,
\eeq
where,
\be
z \equiv \tanh Ht = \left( 1 - H^{-2} a^{-2} \right)^\frac{1}{2} .
\ee

\section{Adiabatic solutions}

The general solution of the function $\sigma$ in (\ref{SIGm01}) is given by (\ref{SIGm02}) with $\sigma_1$ and $\sigma_2$ satisfying (\ref{SIGm03}). Let us not consider the following two WKB solutions of (\ref{SIGm03}) satisfying the given boundary condition  
\be
\sigma_1(t)  = \sqrt{\frac{k}{M \omega_k}} \cos S \ , \ \sigma_2(t) = \sqrt{\frac{k}{M \omega_k}} \sin S ,
\ee
where, $M(t) = a^3(t)$, $\omega_k(t)$ is given by (\ref{OM01}), and
\be
S(t) = \int^t \omega(t') dt' .
\ee
Thus,
\be\label{ADL01}
\sigma \equiv  \sigma_k = \sqrt{\frac{k}{M \omega_k}}  \ , \ \tau_k(t) = \frac{1}{k} \int^t\omega(t') dt'   ,
\ee
which satisfy the asymptotic conditions
\be
\sigma \rightarrow \frac{1}{a} \ , \ \tau_k \rightarrow \eta , 
\ee
in the limit, $\frac{k}{a} \rightarrow \infty$, for which, $\omega_k \rightarrow \frac{k}{a}$ and $\zeta_\textbf{k}(\tau_k) \rightarrow \chi_\textbf{k}(\eta)$. The function $\sigma$ given by (\ref{ADL01}) satisfies (\ref{SIGm01}) provided that
\be
\sigma \left( \frac{\dot{M}^2}{4 M^2} - \frac{\ddot{M}}{2 M} + \frac{3 \dot{\omega}^2}{4 \omega^2}  - \frac{\ddot{\omega}}{2 \omega} \right) \rightarrow 0 ,
\ee
in some appropriate limit. In the case of minimal coupling this is accomplished whenever
\be\label{LIM02}
-\frac{\sigma}{2} \left( \frac{(3\alpha + 2)^2 - 6\alpha^2 }{(1+\alpha)^2} \frac{\dot{a}^2}{2 a^2}  + \frac{3\alpha + 2}{1+\alpha} \frac{\ddot{a}}{a} \right) \rightarrow 0 ,
\ee
where
\be
\alpha \equiv \frac{m^2 a^2 }{k^2} .
\ee
The limit (\ref{LIM02}) is satisfied for large values of the physical modes, $\frac{k}{a} \gg 1$, for which $\sigma \rightarrow \frac{1}{a}$ and $\alpha\rightarrow 0$, provided that
\be
\frac{1}{a} \left( \frac{\dot{a}^2}{a^2} + \frac{\ddot{a}}{a} \right) \rightarrow 0 .
\ee
It is also satisfied for large values of the scale factor, $\frac{k}{a} \ll 1$, for which $\sigma \rightarrow \sqrt{\frac{k}{a^3 m}}$ and $\alpha \gg 1$, whenever
\be
\sqrt{\frac{k}{m a}} \frac{1}{a} \left( \frac{9 \dot{a}^2}{4 a^2} + \frac{3 \ddot{a}}{2 a} \right) \rightarrow 0 .
\ee
Therefore, the adiabatic solution (\ref{ADL01}) is valid for many cases of interest, including those for which
\be
\frac{\dot{a}^2}{a^3} \rightarrow 0 \ \text{  and, } \ \frac{\ddot{a}}{a^2} \rightarrow 0 .
\ee
In those cases, the value of $\nu$ in (\ref{NU02}) can be approximated by
\be
\nu(t) = \frac{i}{4 \omega} \left( \frac{\dot{M}}{M} + \frac{\dot{\omega}}{\omega} \right) = \frac{i}{4 \omega} \left( \frac{2+3\alpha}{1+\alpha}\right)  \frac{\dot{a}}{a} \rightarrow \frac{i}{2} \frac{\dot{a}}{k} ,
\ee
in the limit of large modes, which  is similar to that given in (\ref{Nflat}) and in (\ref{NUkcds}). The energy density\footnote{Above the zero point energy density.} associated to the mode $k$ is given, in the limit $ m \ll \frac{k}{a}$, by
\be
\varepsilon_k = \frac{\dot{a}^2}{4 a^4} \frac{1}{k} ,
\ee
so the energy density of entanglement is  given by
\be
\varepsilon \propto \int_0^{k_\text{m}} dk \ k^2 \varepsilon_k   \approx \frac{k_\text{m}^2}{4 a^2} \frac{\dot{a}^2}{a^2},
\ee
where an ultraviolet cut-off,  $k_\text{m}$,  has been introduced. The dynamics of the background spacetime turns out to be then modified by the existence of the energy of entanglement between the particles and antiparticles of the invariant vacuum. The modified Friedmann equation would read, in the region $ a \gg k_\text{m}$,
\be
\left( \frac{\dot{a}}{a} \right)^2 \approx \rho_0 \left( 1 +  \frac{k_\text{m}^2}{4 a^2} \right) .
\ee
For instance, let us consider a flat deSitter universe for which $\rho_0 = \Lambda \equiv H^2$. The Friedmann equation would be modified by the existence of the entanglement between the modes of the particles-antiparticle pairs, and the scale factor would end up evolving like
\be
a(t) \approx \frac{k_\text{m}}{ 2} \sinh H\Delta t ,
\ee
instead of the customary exponential expansion (\ref{SF01fds}). Thus, the entanglement between the modes of the scalar field would produce a departure from the evolution of the initial flat deSitter spacetime that might be observable.

Finally, let us notice that the amplitude of fluctuations (\ref{AF01fds}) turns out to be given, in the case of the adiabatic solution (\ref{ADL01}), by
\be
\delta\phi(k_\text{ph}) \propto \frac{k^\frac{3}{2}}{2 \pi M^\frac{1}{2} \omega^\frac{1}{2}}  \approx \left\{
	       \begin{array}{ll}
		 \frac{k_\text{ph}}{2 \pi}  ,    &  k_\text{ph} \gg m , \\
		  &  \\
		 \frac{m}{2\pi} \left( \frac{k_\text{ph}}{m} \right)^\frac{3}{2}  ,   &  k_\text{ph} \ll m ,
	       \end{array}
	     \right.
\ee
which are scale independent in both sub-curvature and super-curvature scales. Besides, for short-wavelength modes the spectrum is in agreement with the spectrum of fluctuations in Minkowski spacetime \cite{Mukhanov2007} and thus the field modes are not significantly  affected on sub-curvature scales, as it was expected.

\section{Conclusions}

We have applied the method originally developed by Lewis and Riesenfeld and further developed by others for obtaining the invariant representation of the field modes of a charged scalar field. Then, we have assumed that the field is in the vacuum state of the invariant representation, in terms of which there is no particle production at all at any time, because it represents the ground state along the entire evolution of the field. In order to fix the vacuum state we have further imposed that the largest modes of the field must not feel the curvature of the spacetime, which is a boundary condition ultimately rooted in the equivalence principle of the theory of relativity.

We have assumed however that the observable modes of the field are those described by the instantaneously diagonal representation of the Hamiltonian at a given moment of time, when the observer is performing the measurement. In a small local region of the space, any practical particle detector would measure in practice no particles at all. However, on cosmological grounds it turns out that the vacuum state of the invariant representation is full of particle-antiparticle pairs of the diagonal representation, which are created with opposite momentum in entangled states. The quantum state of each single component is given by a quasi-thermal state with a specific temperature of entanglement that measures the rate of entanglement between the component of the created pair.

We have computed the thermodynamical magnitudes of entanglement and represented the energy density of entanglement in the case of a DeSitter spacetime. It is large for the early phases of the universes and becomes very small for an evolved universe like ours.

We have also computed the vacuum state of the invariant representation for a charged scalar field that propagates in a homogeneous and isotropic spacetime that evolves adiabatically. The energy density of the particles of the field modifies the Friedmann equation producing a departure from the unperturbed evolution that could be detected,at least in principle.

We have computed the amplitude of the vacuum fluctuations. In the case of a DeSitter spacetime the amplitude of fluctuations are the expected one. For a general spacetime that evolves adiabatically, they become scale independent for both sub-curvatures and super-curvatures scales.

This work supplies us with a new point of view for the evolution of matter and radiation fields in curved spacetime that can help us to make new further developments, particularly in the context of the thermodynamics of entanglement in curved spacetime backgrounds.

\section*{Acknowledgments}


\bibliographystyle{apsrev4-1}

%

\end{document}